\documentclass[aps,prx,twocolumn,superscriptaddress,showpacs,amsmath,amssymb,notitlepage,longbibliography]{revtex4-1}

\usepackage{amsfonts}
\usepackage{amssymb}
\usepackage{graphicx}
\usepackage{color}

\usepackage{bm}
\usepackage{braket}

\usepackage{hyperref}

\usepackage[normalem]{ulem}

\begin{document}
\title{The Bloch theorem in the presence of an additional conserved charge}
\author{Haruki Watanabe}\email{hwatanabe@g.ecc.u-tokyo.ac.jp}
\affiliation{Department of Applied Physics, University of Tokyo, Tokyo 113-8656, Japan}
\begin{abstract}
The Bloch theorem is a general theorem restricting the persistent current associated with a conserved U(1) charge in a ground state or in a thermal equilibrium. 
It gives an upper bound of the magnitude of the current density, which is inversely proportional to the system size. 
In a recent paper, Else and Senthil applied the argument for the Bloch theorem to a generalized Gibbs ensemble, assuming the presence of an additional conserved charge, and predicted a nonzero current density in the nonthermal steady state [D. V. Else and T. Senthil, Phys. Rev. B 104, 205132 (2021)]. 
In this work, we provide a complementary derivation based on the canonical ensemble, given that the additional charge is strictly conserved within the system by itself.  
Furthermore, using the example where the additional conserved charge is the momentum operator, we discuss that the persistent current tends to vanish  when the system is in contact with an external momentum reservoir in the co-moving frame of the reservoir.
\end{abstract}
\maketitle

\section{Introduction}
The Bloch theorem is a fundamental theorem stating that the current density of a conserved U(1) charge vanishes in thermodynamically large systems~\cite{PhysRev.75.502,doi:10.1143/JPSJ.65.3254,PhysRevD.92.085011,Watanabe2019,Tada2016,Bachmann2020}.  The theorem applies quite generally regardless of the detailed form of the Hamiltonian and the presence or absence of many-body interactions and impurities.  The only assumptions of the theorem are the locality of the Hamiltonian and the U(1) symmetry that defines the conserved charge~\cite{Watanabe2019}.
Although the theorem was originally derived for a ground state~\cite{PhysRev.75.502}, it has been generalized to a thermal equilibrium described by a canonical ensemble and a grand canonical ensemble~\cite{doi:10.1143/JPSJ.65.3254,Watanabe2019,Tada2016}.  

Recently, there appeared an interesting proposal~\cite{ElseSenthil} which, among other things, found that the proof of the Bloch theorem can be used to predict a nonzero persistent current in systems described by a ``generalized Gibbs ensemble."  Although the generalized Gibbs ensemble usually describes the quench dynamics of isolated systems with an extensive number of conserved charges~\cite{Vidmar_2016,Ueda2020}, here the system has only a few number of conserved quantities: the Hamiltonian $\hat{H}$, the U(1) charge $\hat{Q}$ that defines the current density, and the additional charge $\hat{\Gamma}$.  However, if the system is completely isolated and the additional charge is strictly conserved within the system, it is more natural not to take the ensemble average for the charge. However, since the derivation in Ref.~\cite{ElseSenthil} is specialized to the generalized Gibbs ensemble, it is not clear if the same conclusion can be derived within this picture. Moreover, since condensed matter systems are usually in contact with an environment, we may understand the generalized Gibbs ensemble as a result of exchanges of conserved quantities with reservoirs. 
For example, when $\hat{\Gamma}$ is the momentum operator, a nonzero persistent current implies a relative motion between the system and the reservoir, and whether such a motion persists after equilibration needs to be clarified.

We address these issues in this work.  We develop a complementary approach based on a grand canonical ensemble, assuming the conservation of the additional charge strictly within the focused system, and derive an equivalent conclusion in this setting. We then discuss the case when the additional conserved charge is the momentum operator associated with the continuous translation symmetry. When the system is in contact with an external momentum reservoir, we find that the velocities of the system and the reservoir must coincide in the generalized Gibbs ensemble, implying that the persistent current density vanishes in the laboratory frame where the reservoir is assumed to be stationary.  In contrast, when the system by itself is playing the role of the reservoir for its subsystem~\cite{Vidmar_2016},  this argument simply implies a uniform flow over the entire system. 

\section{Review of the Bloch theorem in Gibbs ensemble}
\label{sec:review}
In this section, we review the standard discussions for the Bloch theorem for a thermal equilibrium~\cite{doi:10.1143/JPSJ.65.3254,Watanabe2019,Tada2016}  to set a basis for later sections.
\subsection{Setting}
Let us consider a one-dimensional system described by a Hamiltonian $\hat{H}=\int_0^Ldx\hat{h}_x$ under the periodic boundary condition. 
Here we assume continuum models having in mind the example of momentum conservation discussed in Sec.~\ref{sec:difficulties}. However, lattice models can be treated in the same way as we give more details in Sec.~\ref{sec:energy}.  The Hamiltonian is invariant under a U(1) symmetry that defines a conserved charge $\hat{Q}=\int_0^Ldx\hat{n}_x$ where $\hat{n}_x$ is the charge density operator. We assume $[\hat{h}_x,\hat{Q}]=[\hat{n}_x,\hat{n}_{x'}]=0$ for all $x$ and $x'$.  Let $\hat{j}_{x}$  be the corresponding current operator, satisfying the continuity equation
\begin{equation}
i[\hat{H},\hat{n}_x]+\partial_x\hat{j}_{x}=0.
\label{continuity}
\end{equation}
It follows that the expectation value of $\hat{j}_x$ is $x$-independent in any stationary state. Thus the equilibrium current density can be fully characterized by the expectation value of the averaged current density $\hat{j}\equiv L^{-1}\int_0^Ldx\hat{j}_x$. The averaged current density operator can be obtained by introducing the uniform gauge field $A$ associated with the U(1) charge $\hat{Q}$ and by taking a derivative with respect to it:
\begin{align}
\hat{j}&=\frac{1}{L}\frac{\partial\hat{H}(A)}{\partial A}\Big|_{A=0}.
\label{avecurrent}
\end{align}
Taking another derivative, we get
\begin{align}
\hat{\sigma}\equiv\frac{1}{L}\frac{\partial^2\hat{H}(A)}{\partial A^2}\Big|_{A=0},
\label{defsigma}
\end{align}
whose expectation value gives the frequency sum of the optical conductivity~\cite{Resta_2018,WatanabeOshikawa2020}.

\subsection{Canonical ensemble}
\label{sec:C}
When the system is weakly connected to a reservoir that supplies or absorbs only heat, it is natural to restrict the state into the subspace of a particular value of the U(1) charge $\hat{Q}=Q_0\in\mathbb{N}$. The thermal equilibrium is described by the (restricted) canonical ensemble 
\begin{align}
&\hat{\rho}_{\text{c}}^{(Q_0)}\equiv \frac{1}{Z_{\text{c}}}e^{-\beta\hat{H}}\hat{\mathcal{P}}_{\hat{Q}=Q_0},\\
&Z_{\text{c}}^{(Q_0)}\equiv\text{Tr}\left(e^{-\beta\hat{H}}\hat{\mathcal{P}}_{\hat{Q}=Q_0}\right),
\end{align}
where $\beta\equiv T^{-1}$ is the inverse temperature and $\hat{\mathcal{P}}_{\hat{Q}=Q_0}$ is the projector onto the $\hat{Q}=Q_0$ subspace. This is equivalent to say that the trace is over the subspace of $\hat{Q}=Q_0$ only.  

The Bloch theorem states that the equilibrium current density is bounded above by $cL^{-1}$ with a constant $c>0$:
\begin{align}
\text{Tr}(\hat{\rho}_{\text{c}}^{(Q_0)}\hat{j})=O(L^{-1}).
\end{align}
To see this, let us introduce the twist operator 
\begin{align}
\hat{U}_w\equiv e^{i2\pi wL^{-1}\int_{0}^Ldx x\hat{n}_x},
\end{align}
which is a large gauge transformation that changes the gauge field $A$ by $2\pi w/L$ and is consistent with the periodic boundary condition when the winding number $w$ is an integer. From Eqs.~\eqref{avecurrent} and \eqref{defsigma}, we find
\begin{align}
\hat{U}_w^\dagger \hat{H}\hat{U}_w&=\hat{H}(A=\tfrac{2\pi w}{L})\notag\\
&=\hat{H}+2\pi w\hat{j}+\frac{(2\pi w)^2}{2L}\hat{\sigma}+O(L^{-2}).
\label{UdHU}
\end{align}

The Bloch theorem can be most easily proven by contradiction. On one hand, the canonical ensemble $\hat{\rho}_{\text{c}}^{(Q_0)}$ minimizes the free energy
\begin{align}
F_{\text{c}}^{(Q_0)}(\hat{\rho})\equiv\text{Tr}\left(\hat{\mathcal{P}}_{\hat{Q}=Q_0}\big(\hat{\rho}\hat{H}+T\hat{\rho}\log\hat{\rho}\big)\right).
\end{align}
On the other hand, the free energy for the state $\hat{U}_w \hat{\rho}_{\text{c}}^{(Q_0)}\hat{U}_w^\dagger$ reads
\begin{align}
F_{\text{c}}^{(Q_0)}(\hat{U}_w \hat{\rho}_{\text{c}}^{(Q_0)}\hat{U}_w^\dagger)&=F_{\text{c}}^{(Q_0)}(\hat{\rho}_{\text{c}}^{(Q_0)})+2\pi w\text{Tr}(\hat{\rho}_{\text{c}}^{(Q_0)}\hat{j})\notag\\
&\quad+\frac{(2\pi w)^2}{2L}\text{Tr}(\hat{\rho}_{\text{c}}^{(Q_0)}\hat{\sigma})+O(L^{-2}).
\end{align}
In the derivation, we used the fact that
\begin{align}
\hat{U}_w^\dagger \hat{Q}\hat{U}_w=\hat{Q},
\label{UdQU}
\end{align}
which follows from the assumed commutation relation among the charge density operators $\hat{n}_x$.
Since $w$ is an arbitrary integer, $F_{\text{c}}^{(Q_0)}(\hat{U}_w \hat{\rho}_{\text{c}}^{(Q_0)}\hat{U}_w^\dagger)$ can be lower than the free energy of the canonical ensemble $F_{\text{c}}^{(Q_0)}(\hat{\rho}_{\text{c}}^{(Q_0)})=-T\log Z_{\text{c}}^{(Q_0)}$. The contradiction with the variational principle can be avoided only when
\begin{equation}
\big|\text{Tr}(\hat{\rho}_{\text{c}}^{(Q_0)}\hat{j})\big|\leq\frac{\pi}{L}\text{Tr}(\hat{\rho}_{\text{c}}^{(Q_0)}\hat{\sigma})+O(L^{-2}).
\label{Bloch1}
\end{equation}
On the right hand side, $\text{Tr}(\hat{\rho}_{\text{c}}^{(Q_0)}\hat{\sigma})$ is the frequency sum of the optical conductivity~\cite{TakasanOshikawaWatanabe}.
For a later purpose, we remind ourselves that the chemical potential $\mu$ associated with the charge $\hat{Q}$ in this approach is given by 
\begin{align}
\mu=\frac{\partial F_{\text{c}}^{(Q_0)}(\hat{\rho}_{\text{c}}^{(Q_0)})}{\partial Q_0}.
\label{muGC}
\end{align}

\subsection{Grand canonical ensemble}
\label{sec:GC}
Next, let us assume that the reservoir can supply or absorb the U(1) charge $\hat{Q}$ as well. Then we take the grand canonical ensemble
\begin{align}
&\hat{\rho}_{\text{gc}}^{(\mu)}\equiv \frac{1}{Z_{\text{gc}}^{(\mu)}}e^{-\beta(\hat{H}-\mu\hat{Q})},\\
&Z_{\text{gc}}^{(\mu)}\equiv\text{Tr}\,e^{-\beta(\hat{H}-\mu\hat{Q})}.
\end{align}
The relevant free energy in this setting is
\begin{align}
F_{\text{gc}}^{(\mu)}(\hat{\rho})\equiv\text{Tr}\big(\hat{\rho}(\hat{H}-\mu \hat{Q})+T\hat{\rho}\log\hat{\rho}\big).
\label{Fgc}
\end{align}
To make a connection to the canonical ensemble in Sec.~\ref{sec:C}, the chemical potential $\mu$ should be fixed in such a way that the expectation value
\begin{align}
Q_0=\text{Tr}(\hat{\rho}_{\text{gc}}^{(\mu)}\hat{Q})=-\frac{\partial F_{\text{gc}}^{(\mu)}(\hat{\rho}_{\text{gc}}^{(\mu)})}{\partial \mu}
\label{Q0c}
\end{align}
matches the value assumed in the (restricted) canonical ensemble. 
In a thermal equilibrium, the temperatures and the chemical potentials of the system and the reservoir are balanced, and the free energy  $F_{\text{gc}}^{(\mu)}(\hat{\rho}_{\text{gc}}^{(\mu)})=-T\log Z_{\text{gc}}^{(\mu)}$ is a function of the common values of $\beta$ and $\mu$. 

One can repeat the same argument as in Sec.~\ref{sec:C} for the free energy in Eq.~\eqref{Fgc}  to derive the Bloch theorem for the grand canonical ensemble
\begin{equation}
\big|\text{Tr}(\hat{\rho}_{\text{gc}}^{(\mu)}\hat{j})\big|\leq\frac{\pi}{L}\text{Tr}(\hat{\rho}_{\text{gc}}^{(\mu)}\hat{\sigma})+O(L^{-2}).
\label{Bloch2}
\end{equation}
Hence, we obtain the equivalent conclusion regardless of whether we assume the (restricted) canonical ensemble or the grand canonical ensemble, as anticipated from the equivalence of the canonical ensemble and the grand canonical ensemble~\cite{landau}.

\section{Bloch theorems with additional conserved charges}
\label{sec:DS}
In this section, we discuss how the discussions on the Bloch theorem in Sec.~\ref{sec:review} are modified in the presence of an additional conserved charge.
We assume that the system has another conserved charge $\hat{\Gamma}$ that commutes with 
$\hat{Q}$, i.e., $[\hat{\Gamma},\hat{H}]=[\hat{\Gamma},\hat{Q}]=0$. 

\subsection{Generalized Gibbs ensemble}
\label{sec:GG}
Let us first reproduce the discussion in Ref.~\cite{ElseSenthil}. 
Suppose that the reservoir can supply or absorb the additional charge $\hat{\Gamma}$ as well. 
Then we describe the system by the generalized Gibbs ensemble~\cite{ElseSenthil},  
\begin{align}
&\hat{\rho}_{\text{gG}}^{(\mu,\eta)}\equiv \frac{1}{Z_{\text{gG}}^{(\mu,\eta)}}e^{-\beta(\hat{H}-\mu\hat{Q}-\eta\hat{\Gamma})},\\
&Z_{\text{gG}}^{(\mu,\eta)}\equiv\text{Tr}\,e^{-\beta(\hat{H}-\mu\hat{Q}-\eta\hat{\Gamma})}.
\end{align}
If one repeats the above argument for the free energy
\begin{align}
F_{\text{gG}}^{(\mu,\eta)}(\hat{\rho})\equiv\text{Tr}\big(\hat{\rho}(\hat{H}-\mu \hat{Q}-\eta\hat{\Gamma})+T\hat{\rho}\log\hat{\rho}\big),
\label{FGG}
\end{align}
this time one instead finds~\cite{ElseSenthil}
\begin{align}
\text{Tr}\big(\hat{\rho}_{\text{gG}}^{(\mu,\eta)}\hat{j}\big)=\eta\text{Tr}\big(\hat{\rho}_{\text{gG}}^{(\mu,\eta)}\hat{\xi}\big)+O(L^{-1}).
\label{Bloch32}
\end{align}
One can also obtain an inequality similar to Eq.~\eqref{Bloch1} and Eq.~\eqref{Bloch2}:
\begin{align}
\big|\text{Tr}\big(\hat{\rho}_{\text{gG}}^{(\mu,\eta)}(\hat{j}-\eta\hat{\xi})\big)\big|\leq\frac{\pi}{L}\big|\text{Tr}\big(\hat{\rho}_{\text{gG}}^{(\mu,\eta)}(\hat{\sigma}-\eta\hat{\zeta})\big)\big|+O(L^{-2}).
\label{Bloch3}
\end{align}
These results implies that the persistent current density $\text{Tr}\big(\hat{\rho}_{\text{gG}}^{(\mu,\eta)}\hat{j}\big)$ does not vanish in the limit of large $L$ when
$\eta\neq0$ and $\text{Tr}\big(\hat{\rho}_{\text{gG}}^{(\mu,\eta)}\hat{\xi}\big)\neq0$~\cite{ElseSenthil}.  
Here $\hat{\xi}$ and $\hat{\zeta}$ are the contributions from $\hat{\Gamma}$ defined by
\begin{align}
\hat{\xi}&\equiv\frac{1}{L}\frac{\partial\hat{\Gamma}(A)}{\partial A}\Big|_{A=0},\\
\hat{\zeta}&\equiv\frac{1}{L}\frac{\partial^2\hat{\Gamma}(A)}{\partial A^2}\Big|_{A=0},
\end{align}
where $\hat{\Gamma}(A)$ is the conserved charge $\hat{\Gamma}$ under the uniform gauge field $A$ associated with $\hat{Q}$.  In the derivation of Eqs.~\eqref{Bloch32} and \eqref{Bloch3}, we used
\begin{align}
\hat{U}_w^\dagger \hat{\Gamma}\hat{U}_w&=\hat{\Gamma}(A=\tfrac{2\pi w}{L})\notag\\
&=\hat{\Gamma}+2\pi w\hat{\xi}+\frac{(2\pi w)^2}{2L}\hat{\zeta}+O(L^{-2}).
\label{eqGamma}
\end{align}
In examples discussed in Ref.~\cite{ElseSenthil}, $\hat{\zeta}$ and higher order terms in $L^{-1}$ were absent, but we will see a more general case in Sec.~\ref{sec:energy}.

The condition $\text{Tr}\big(\hat{\rho}_{\text{gG}}^{(\mu,\eta)}\hat{\xi}\big)\neq0$ for the nonvanishing persistent current in the large $L$ limit may not be intuitive. Alternatively, we can also write
\begin{align}
&\frac{d}{d\eta}\text{Tr}\big(\hat{\rho}_{\text{gG}}^{(\mu,\eta)}\hat{j}\big)\Big|_{\eta=0}\notag\\
&=\beta\text{Tr}\big(\hat{\rho}_{\text{gc}}^{(\mu)}\hat{\Gamma}\hat{j}\big)-\beta\text{Tr}\big(\hat{\rho}_{\text{gc}}^{(\mu)}\hat{\Gamma}\big)\text{Tr}\big(\hat{\rho}_{\text{gc}}^{(\mu)}\hat{j}\big).\label{overlap}
\end{align}
The second term vanishes in the large $L$ limit as suggested by Eq.~\eqref{Bloch2}. Hence, the operator $\hat{\Gamma}$ should have a finite overlap with the current operator $\hat{j}$ [i.e., $\text{Tr}\big(\hat{\rho}_{\text{gc}}^{(\mu)}\hat{\Gamma}\hat{j}\big)\neq0$] in order to induce the persistent U(1) current in the linear response.

In the generalized Gibbs ensemble,  the ``chemical potential" $\eta$ for $\hat{\Gamma}$ should be set in such a way that the expectation value of the charge $\hat{\Gamma}$, given by
\begin{align}
\Gamma_0=\text{Tr}(\hat{\rho}_{\text{gG}}^{(\mu,\eta)}\hat{\Gamma})=-\frac{\partial F_{\text{gG}}^{(\mu,\eta)}(\hat{\rho}_{\text{gG}}^{(\mu,\eta)})}{\partial \eta},
\end{align}
agrees to the value fixed by the initial condition. This is analogous to the chemical potential $\mu$ in the grand canonical ensemble.

The key behind the possible nonzero current in Eq.~\eqref{Bloch32} is the mismatch between the Hamiltonian $\hat{H}$ that defines the current operator in Eq.~\eqref{avecurrent} and 
\begin{align}
\hat{H}^{(\mu,\eta)}\equiv\hat{H}-\mu\hat{Q}-\eta\hat{\Gamma} 
\end{align}
that appears in the free energy in Eq.~\eqref{FGG}. In other words, the expectation value of the effective current operator corresponding to $\hat{H}^{(\mu,\eta)}$,
\begin{equation}
\hat{j}^{(\mu,\eta)}\equiv\frac{1}{L}\frac{\partial\hat{H}^{(\mu,\eta)}(A)}{\partial A}\Big|_{A=0}=\hat{j}-\eta\hat{\xi},
\end{equation}
is $O(L^{-1})$ in the generalized Gibbs ensemble. A similar mismatch exists for the grand canonical ensemble in Sec.~\ref{sec:GC} but it did not lead to any change because $\hat{U}_w$ and $\hat{Q}$ commute [see Eq.~\eqref{UdQU}].  However, when we interpret the system by itself playing the role of the reservoir for its subsystem, then the entire system may be described by the canonical or the grand canonical ensemble discussed in Sec.~\ref{sec:review}. Then the $-\eta\hat{\Gamma}$ term, which was the key in the above derivation, is missing from the free energy. 
Then the question is whether we can reproduce the same result even in this case.

\subsection{Canonical ensemble with additional conserved charge}
\label{GCACC}
Let us revisit the canonical ensemble but this time paying attention to the presence of the additional conserved charge $\hat{\Gamma}$.
We assume that the reservoir supplies or absorbs the heat only, and  $\hat{Q}$ and $\hat{\Gamma}$ are strictly conserved.
Then we restrict the state into the subspace of $\hat{Q}=Q_0$ and $\hat{\Gamma}=\Gamma_0$, where $\Gamma_0$ here is an eigenvalue of $\hat{\Gamma}$. We thus consider  the (restricted) canonical ensemble
\begin{align}
&\hat{\rho}_{\text{c}}^{(Q_0,\Gamma_0)}\equiv \frac{1}{Z_{\text{c}}^{(Q_0,\Gamma_0)}}e^{-\beta\hat{H}}\hat{\mathcal{P}}_{\hat{Q}=Q_0}\hat{\mathcal{P}}_{\hat{\Gamma}=\Gamma_0},\\
&Z_{\text{c}}^{(Q_0,\Gamma_0)}\equiv\text{Tr}\left(e^{-\beta\hat{H}}\hat{\mathcal{P}}_{\hat{Q}=Q_0}\hat{\mathcal{P}}_{\hat{\Gamma}=\Gamma_0}\right).
\end{align}
Note that, in general, the state $\hat{\rho}_{\text{c}}^{(Q_0,\Gamma_0)}$ does not remain within the $\hat{\Gamma}=\Gamma_0$ subspace under the action of the twist operator, because  Eq.~\eqref{eqGamma} implies
\begin{align}
\Gamma_w&\equiv\text{Tr}\big((\hat{U}_w \hat{\rho}_{\text{c}}^{(Q_0,\Gamma_0)}\hat{U}_w^\dagger)\hat{\Gamma}\big)=\text{Tr}\big(\hat{\rho}_{\text{c}}^{(Q_0,\Gamma_0)}(\hat{U}_w^\dagger\hat{\Gamma}\hat{U}_w)\big)\notag\\
&=\Gamma_0+2\pi w\text{Tr}\big(\hat{\rho}_{\text{c}}^{(Q_0,\Gamma_0)}\hat{\xi}\big)+O(L^{-1}).
\label{Gamma0p}
\end{align}
Therefore, although it is still true that $\hat{\rho}_{\text{c}}^{(Q_0,\Gamma_0)}$ minimizes the free energy
\begin{align}
F_{\text{c}}^{(Q_0,\Gamma_0)}(\hat{\rho})\equiv\text{Tr}\left(\hat{\mathcal{P}}_{\hat{Q}=Q_0}\hat{\mathcal{P}}_{\hat{\Gamma}=\Gamma_0}\big(\hat{\rho}\hat{H}+T\hat{\rho}\log\hat{\rho}\big)\right),
\end{align}
$\hat{U}_w \hat{\rho}_{\text{c}}^{(Q_0,\Gamma_0)}\hat{U}_w^\dagger$ does not provide a state with lower free energy within the $\hat{\Gamma}=\Gamma_0$ subspace.
This is why the proof of the Bloch theorem in Sec.~\ref{sec:review} is not applicable to $\hat{\rho}_{\text{c}}^{(Q_0,\Gamma_0)}$ when $\Gamma_w\neq\Gamma_0$.

Now, let us demonstrate
\begin{align}
\text{Tr}\big(\hat{\rho}_{\text{c}}^{(Q_0,\Gamma_0)}\hat{j}\big)=\eta\text{Tr}\big(\hat{\rho}_{\text{c}}^{(Q_0,\Gamma_0)}\hat{\xi}\big)+O(L^{-1})
\label{Bloch4}
\end{align}
in the restricted canonical ensemble $\hat{\rho}_{\text{c}}^{(Q_0,\Gamma_0)}$. 
In this picture, the potential $\eta$ is given by
\begin{align}
\eta=\frac{\partial F_{\text{c}}^{(Q_0,\Gamma_0)}(\hat{\rho}_{\text{c}}^{(Q_0,\Gamma_0)})}{\partial \Gamma_0},
\label{etaRGC}
\end{align}
just like the chemical potential $\mu$ in the canonical ensemble in Sec.~\ref{sec:C} [see.~Eq.~\eqref{muGC}].

To proceed, here we assume that $\hat{\xi}$ in Eq.~\eqref{eqGamma} is given by a function of $\hat{Q}$, i.e., 
\begin{equation}
\hat{\xi}=\xi(\hat{Q})
\end{equation}
and that $\hat{\zeta}$ and all higher terms vanish. This was the case for the momentum operator, for which $\hat{\xi}=\hat{Q}/L$, and also for emergent symmetries, for which $\hat{\xi}$ is a constant $m/(2\pi)$ with $m\in\mathbb{Z}$~\cite{ElseSenthil}. Then $\Gamma_w$ in Eq.~\eqref{Gamma0p} becomes $\Gamma_w=\Gamma_0+2\pi w\xi(Q_0)$, which should be an eigenvalue of $\hat{\Gamma}$.  We are interested in the case when $\xi(Q_0)\neq0$.

To evaluate the change of the free energy, we define
\begin{align}
&(\Delta F_{\text{c}}^{(Q_0,\Gamma_0)})_w\equiv F_{\text{c}}^{(Q_0,\Gamma_w)}(\hat{\rho}_{\text{c}}^{(Q_0,\Gamma_w)})-F_{\text{c}}^{(Q_0,\Gamma_0)}(\hat{\rho}_{\text{c}}^{(Q_0,\Gamma_0)})\\
&(\Delta \Gamma)_w\equiv\Gamma_w-\Gamma_0=2\pi w\xi(Q_0).
\end{align}
Based on the variational principle, we find
\begin{align}
&(\Delta F_{\text{c}}^{(Q_0,\Gamma_0)})_w\notag\\
&\leq F_{\text{c}}^{(Q_0,\Gamma_w)}(\hat{U}_w \hat{\rho}_{\text{c}}^{(Q_0,\Gamma_0)}\hat{U}_w^\dagger)-F_{\text{c}}^{(Q_0,\Gamma_0)}(\hat{\rho}_{\text{c}}^{(Q_0,\Gamma_0)})\notag\\
&=2\pi w\text{Tr}\left(\hat{\rho}_{\text{c}}^{(Q_0,\Gamma_0)}\hat{j}\right)+O(L^{-1}),
\end{align}
where we used Eq.~\eqref{UdHU} in the last step. Thus, assuming $\xi(Q_0)>0$ and neglecting $O(L^{-1})$ corrections, we find
\begin{align}
\frac{(\Delta F_{\text{c}}^{(Q_0,\Gamma_0)})_{+1}}{(\Delta \Gamma)_{+1}}\leq \frac{\text{Tr}\left(\hat{\rho}_{\text{c}}^{(Q_0,\Gamma_0)}\hat{j}\right)}{\xi(Q_0)}\leq \frac{(\Delta F_{\text{c}}^{(Q_0,\Gamma_0)})_{-1}}{(\Delta \Gamma)_{-1}}.\label{ineq1}
\end{align}
When the free energy density $F_{\text{c}}^{(Q_0,\Gamma_0)}(\hat{\rho}_{\text{c}}^{(Q_0,\Gamma_0)})/L$ is a smooth function of $\Gamma_0/L$ in the limit of large $L$, the left-most side and the right-most side of  \eqref{ineq1} should be identical to $\eta$ in Eq.~\eqref{etaRGC}. Hence we obtain Eq.~\eqref{Bloch4}.
When $\xi(Q_0)<0$, these inequalities are flipped but the conclusion is unchanged. 

The above derivation suggests that, when $\Gamma_0$ is allowed to vary, it will be spontaneously tuned to the value that minimizes the free energy $F_{\text{c}}^{(Q_0,\Gamma_0)}(\hat{\rho}_{\text{c}}^{(Q_0,\Gamma_0)})=-T\log Z_{\text{c}}^{(Q_0,\Gamma_0)}$. At the minimum of the free energy, $\eta$ in Eq.~\eqref{etaRGC}, hence the persistent current density, vanishes.
In other words, states with nonzero $\eta$ are at most metastable.

\section{Examples}
In this section, we discuss concrete realizations of nonzero persistent current density.
As discussed in Ref.~\cite{ElseSenthil}, $\hat{\xi}$ vanishes for any genuine internal symmetries. Hence, we should explore space-time symmetries or emergent symmetries at a low-energy effective theory.
Since the case of emergent symmetries has been discussed in detail in Ref.~\cite{ElseSenthil}, here we focus on the case where $\hat{\Gamma}$ is associated with a space-time symmetry.

\subsection{Momentum conservation}
\label{sec:difficulties}
Let us discuss the case when $\hat{\Gamma}$ is the momentum operator $\hat{P}$ defined as the generator of the continuous translation:
\begin{align}
e^{-i\epsilon\hat{P}}\hat{n}_{x}e^{i\epsilon\hat{P}}=\hat{n}_{x+\epsilon}.
\end{align}
Following the proof of the Lieb-Schultz-Mattis theorem~\cite{LIEB1961407}, one can show that
\begin{align}
e^{-i\epsilon\hat{P}}\hat{U}_we^{i\epsilon\hat{P}}=\hat{U}_we^{-i \epsilon\,2\pi w\hat{Q}/L},
\end{align}
which implies that
\begin{align}
\hat{U}_w^\dagger\hat{P}\hat{U}_w=\hat{P}+2\pi w\frac{\hat{Q}}{L}.
\end{align}
In the derivation, we used the fact that $\int_{0}^{\epsilon}dx \hat{n}_{x}$ is integer valued.
Thus $\hat{\xi}$ is given by the charge density operator $\hat{Q}/L$ and $\hat{\zeta}$ vanishes~\cite{ElseSenthil}. The chemical potential $\eta$ in this case is the velocity $v$~\cite{ElseSenthil}, which can be understood by recalling that the Legendre transformation between the Lagrangian $L(x,v)$ and the Hamiltonian $H(x,p)$ is given by $H(x,p)=vp-L(x,v)$ in analytical mechanics.

In the grand canonical ensemble discussed in Sec.~\ref{sec:GC}, the temperatures and the chemical potentials of the system and the reservoir must be balanced in the equilibrium. In the same way, in the generalized Gibbs ensemble in Sec.~\ref{sec:GG}, where the system and the reservoir exchange the momentum, the velocity $v_s$ of the system and the velocity $v_r$ of the reservoir must coincide in the steady state. Thus, when the reservoir is an external system, and as far as we observe physical quantities in the laboratory frame in which the reservoir is stationary, $\eta=v_s=v_r=0$. In other words, if we model reservoir's free energy associated with the center of the mass motion as 
\begin{align}
F_r(P_r)=\frac{P_r^2}{2M_r},
\end{align}
the velocity $v_r=\partial F_r(P_r)/\partial P_r=P_r/M_r$ vanishes unless the momentum $P_r$ of the reservoir scales with the total mass of the reservoir $M_r$. 
Then, $\eta$ in Eq.~\eqref{Bloch3} can be set $0$ and the original statement of the Bloch theorem as in Eqs.~\eqref{Bloch1} and \eqref{Bloch2} is recovered. 

On the other hand, if the momentum of the system is absolutely conserved, i.e., when the system and the reservoir do not exchange the momentum, the entire system can be described by the restricted Gibbs ensemble discussed in \ref{GCACC}.
Then, in principle, Eq.~\eqref{Bloch4} with nonzero $\eta$ $(=v_s)$ is allowed. In this setting, the velocity $v_s$ is determined by Eq.~\eqref{etaRGC}, i.e., the derivative of the free energy with respect to the momentum of the system. The well-known examples are (classical) perfect fluids~\cite{PhysRevLett.112.100602} and (quantum) superfluids~\cite{RevModPhys.71.S318}, which lack the viscosity and support a persistent flow.

\subsection{XXZ spin chain}
\label{sec:energy}
Next, let us discuss the persistent U(1) current and the persistent energy current in the spin-1/2 XXZ chain. The Hamiltonian reads $\hat{H}=\sum_{n=1}^L\hat{h}_{n}$ with 
\begin{equation}
\hat{h} _{n}=J(\hat{s}_{n+1}^x\hat{s}_{n}^x+\hat{s}_{n+1}^y\hat{s}_{n}^y+\Delta\hat{s}_{n+1}^z\hat{s}_{n}^z).
\end{equation}
Here, $\hat{s}_{n}^\alpha$ ($\alpha=x,y,z$) is the spin-1/2 operator at the site $n$. We impose the periodic boundary condition and identify $\hat{s}_{n+L}^\alpha$ with $\hat{s}_{n}^\alpha$.

The model has a U(1) charge $\hat{S}^z\equiv\sum_{n=1}^L\hat{s}_{n}^z$. The current operator associated with the link $\bar{n}$ between the sites $n$ and $n+1$ is given by
\begin{align}
\hat{j}_{\bar{n}}^S=J(\hat{s}^{y}_{n+1}\hat{s}^{x}_{n}-\hat{s}^{x}_{n+1}\hat{s}^{y}_{n}),
\end{align}
which satisfies the lattice version of the continuity equation 
\begin{equation}
i[\hat{H},\hat{n}_{n}]+\hat{j}_{\bar{n}}^S-\hat{j}_{\bar{n}-1}^S=0.
\end{equation}
One can also define the energy current operator
\begin{align}
\hat{j}_{\bar{n}}^E&=i[\hat{h}_{n},\hat{h}_{n+1}]\notag\\
&=J^2(\hat{s}^{x}_{n+2}\hat{s}^{z}_{n+1}\hat{s}^{y}_{n}-\hat{s}^{y}_{n+2}\hat{s}^{z}_{n+1}\hat{s}^{x}_{n})\notag\\
&\quad-J^2\Delta(\hat{s}^{x}_{n+2}\hat{s}^{y}_{n+1}-\hat{s}^{y}_{n+2}\hat{s}^{x}_{n+1})\hat{s}^{z}_{n}\notag\\
&\quad-J^2\Delta \hat{s}^{z}_{n+2}(\hat{s}^{x}_{n+1}\hat{s}^{y}_{n}-\hat{s}^{y}_{n+1}\hat{s}^{x}_{n})
\end{align}
satisfying
\begin{equation}
i[\hat{H},\hat{h}_{n}]+\hat{j}_{\bar{n}}^E-\hat{j}_{\bar{n}-1}^E=0.
\end{equation}
For the XXZ model, it is known that the total energy current operator $\hat{\Gamma}\equiv\sum_{n=1}^L\hat{j}_{\bar{n}}^E$ commutes with $\hat{H}$ and $\hat{S}^z$~\cite{PhysRevB.55.11029,10.21468/SciPostPhys.6.1.005}. For this $\hat{\Gamma}$, $\hat{\xi}$ and $\hat{\zeta}$ can also be found from a straightforward calculation:
\begin{align}
\hat{\xi}&=\frac{1}{L}\sum_{n=1}^L2J(\hat{s}_{n+2}^x\hat{s}_{n+1}^z\hat{s}_{n}^x+\hat{s}_{n+2}^y\hat{s}_{n+1}^z\hat{s}_{n}^y)\notag\\
&\quad-\frac{1}{L}\sum_{n=1}^LJ\Delta(\hat{s}_{n+2}^x\hat{s}_{n+1}^x+\hat{s}_{n+2}^y\hat{s}_{n+1}^y)\hat{s}_{n}^z\notag\\
&\quad-\frac{1}{L}\sum_{n=1}^LJ\Delta\hat{s}_{n+2}^z(\hat{s}_{n+1}^x\hat{s}_{n}^x+\hat{s}_{n+1}^y\hat{s}_{n}^y)
\end{align}
and
\begin{align}
\hat{\zeta}&=-\frac{1}{L}\sum_{n=1}^L4J^2(\hat{s}^{x}_{n+2}\hat{s}^{z}_{n+1}\hat{s}^{y}_{n}-\hat{s}^{y}_{n+2}\hat{s}^{z}_{n+1}\hat{s}^{x}_{n})\notag\\
&\quad+\frac{1}{L}\sum_{n=1}^LJ^2\Delta(\hat{s}^{x}_{n+2}\hat{s}^{y}_{n+1}-\hat{s}^{y}_{n+2}\hat{s}^{x}_{n+1})\hat{s}^{z}_{n}\notag\\
&\quad+\frac{1}{L}\sum_{n=1}^LJ^2\Delta \hat{s}^{z}_{n+2}(\hat{s}^{x}_{n+1}\hat{s}^{y}_{n}-\hat{s}^{y}_{n+1}\hat{s}^{x}_{n}).
\end{align}

Let us assume the generalized Gibbs ensemble $\hat{\rho}_{\text{gG}}^{(\mu,\eta)}$ discussed in Sec.~\eqref{sec:GG}.
We are interested in the ground state expectation value of the charge density and the current densities:
\begin{align}
&\langle\hat{s}^z\rangle^{(\mu,\eta)}\equiv \frac{1}{L}\text{tr}[\hat{\rho}_{\text{gG}}^{(\mu,\eta)}\hat{S}^z],\\
&\langle\hat{j}^S\rangle^{(\mu,\eta)}\equiv\text{tr}[\hat{\rho}_{\text{gG}}^{(\mu,\eta)}\hat{j}_{\bar{n}}^S],\\
&\langle\hat{j}^E\rangle^{(\mu,\eta)}\equiv\text{tr}[\hat{\rho}_{\text{gG}}^{(\mu,\eta)}\hat{j}_{\bar{n}}^E].
\end{align}
Their behaviors are restricted by discrete symmetries of the XXZ model: (i) The Hamiltonian $\hat{H}$ and the energy current $\hat{\Gamma}$ are even under the spin flip symmetry $\hat{V}^x\equiv\prod_{n=1}^L\exp[i\pi\hat{s}_{n}^x]$ (i.e., $\hat{H}\hat{V}^x=\hat{V}^x\hat{H}$ and $\hat{\Gamma}\hat{V}^x=\hat{V}^x\hat{\Gamma}$), while the spin $\hat{S}^z$ and hence the spin current $\hat{j}_{\bar{n}}^S$ are odd (i.e., $\hat{S}^z\hat{V}^x=-\hat{V}^x\hat{S}^z$ and $\hat{j}_{\bar{n}}^S\hat{V}^x=-\hat{V}^x\hat{j}_{\bar{n}}^S$). (ii) The Hamiltonian $\hat{H}$ and the spin current $\hat{j}_{\bar{n}}^S$ are even under the time-reversal symmetry symmetry $\hat{\mathcal{T}}\equiv \mathcal{K}\prod_{n=1}^L\exp[i\pi\hat{s}_{n}^y]$ ($\mathcal{K}$ is the complex conjugation), while the spin $\hat{S}^z$ and the energy current $\hat{\Gamma}$ are odd. As a consequence, the spin density and the persistent U(1) current is odd under $\mu$, while the persistent U(1) current and the persistent energy current are odd under $\eta$:
\begin{align}
&\langle\hat{s}^z\rangle^{(\mu,\eta)}=-\langle\hat{s}^z\rangle^{(-\mu,\eta)}=\langle\hat{s}^z\rangle^{(\mu,-\eta)},\label{c1}\\
&\langle\hat{j}^S\rangle^{(\mu,\eta)}=-\langle\hat{j}^S\rangle^{(-\mu,\eta)}=-\langle\hat{j}^S\rangle^{(\mu,-\eta)},\\
&\langle\hat{j}^E\rangle^{(\mu,\eta)}=\langle\hat{j}^E\rangle^{(-\mu,\eta)}=-\langle\hat{j}^E\rangle^{(\mu,-\eta)}.\label{c3}
\end{align}
The operator $\hat{\xi}$ and $\hat{\zeta}$ transform in the same way as $\hat{S}^z$ and $\hat{\Gamma}$, respectively.  From Eqs.~\eqref{c1}--\eqref{c3}, we see that both the spin current and the energy current vanishes when $\eta=0$. This is expected because the generalized Gibbs ensemble $\hat{\rho}_{\text{gG}}^{(\mu,\eta)}$ reduces to the ground canonical ensemble $\hat{\rho}_{\text{gc}}^{(\mu)}$ for which the original version of the Bloch theorem holds as discussed in Sec.~\ref{sec:GC}. 
Furthermore, when $\mu=0$, the generalized Gibbs ensemble $\hat{\rho}_{\text{gG}}^{(0,\eta)}$ is invariant under the spin flip symmetry $\hat{V}^x$, under which $\hat{j}_{\bar{n}}^S$ is odd and $\hat{\Gamma}$ is even. Therefore, $\hat{j}_{\bar{n}}^S$ does not overlap with $\hat{\Gamma}$ and the persistent current is not induced when $\mu=0$ [see Eq.~\eqref{overlap}].
On the other hand, for a generic $\mu$ and $\eta$, we expect that $\langle\hat{j}^S\rangle^{(\mu,\eta)}\neq0$ and $\langle\hat{j}^E\rangle^{(\mu,\eta)}\neq0$. A Bloch-type theorem for the energy current density was recently proven in Refs.~\cite{PhysRevLett.123.060601,Watanabe2}, which states that the persistent energy current vanishes in a ground state or in a thermal equilibrium state in the large $L$ limit. Therefore, the nonvanishing energy current density confirms that the generalized Gibbs ensemble is not a thermal equilibrium but a nonequilibrium steady state.

\subsection{Tight-binding model}
To provide more evidence on nonvanishing values of $\langle\hat{j}^S\rangle^{(\mu,\eta)}$ and $\langle\hat{j}^E\rangle^{(\mu,\eta)}$ in the generalized Gibbs ensemble $\hat{\rho}_{\text{gG}}^{(\mu,\eta)}$, let us examine the noninteracting limit ($\Delta=0$) of the XXZ model in more detail.  In this case the XXZ model can be mapped to the tight-binding model of spinless fermions by the Jordan-Wigner transformation 
\begin{align}
&\hat{s}_{n}^+=\hat{s}_{n}^x+i\hat{s}_{n}^y=e^{+i\pi\sum_{m=1}^{n-1}\hat{c}_{m}^\dagger \hat{c}_{m}}\hat{c}_{n}^\dagger,\\
&\hat{s}_{n}^-=\hat{s}_{n}^x-i\hat{s}_{n}^y=e^{-i\pi\sum_{m=1}^{n-1}\hat{c}_{m}^\dagger \hat{c}_{m}}\hat{c}_{n},\\
&\hat{s}_{n}^z=\hat{c}_{n}^\dagger \hat{c}_{n}-\tfrac{1}{2},
\end{align}
where $\hat{c}_{n}$ is the annihilation operator of the fermion at the site $n$. The boundary condition of the tight-binding model depends on the total number of fermions $\hat{N}=\sum_{n=1}^L\hat{c}_{n}^\dagger \hat{c}_{n}$ in the system; the boundary condition becomes anti-periodic in the even $\hat{N}=N$ sector, while it remains periodic in the odd $\hat{N}=N$ sector~\cite{LIEB1961407}. When $N$ is even, 
we perform a gauge transformation $\hat{U}=\exp(-i\theta_N \sum_{n=1}^Ln\hat{c}_{n}^\dagger \hat{c}_{n})$ with $\theta_N=\pi/L$ to recover the periodic boundary condition. When $N$ is odd, we set $\theta_N=0$. Then we find
\begin{align}
\hat{H}&=\sum_{n=1}^{L}t(e^{-i\theta_N}\hat{c}_{n+1}^\dagger \hat{c}_{n}+e^{i\theta_N}\hat{c}_{n}^\dagger \hat{c}_{n+1}),\\
\hat{S}^z&=\sum_{n=1}^{L}(\hat{c}_{n}^\dagger \hat{c}_{n}-\tfrac{1}{2}),\\
\hat{j}_{\bar{n}}^E&=-it^2(e^{-2i\theta_N}\hat{c}^\dagger_{n+2}\hat{c}_{n}-e^{2i\theta_N}\hat{c}^\dagger_{n}\hat{c}_{n+2}),\\
\hat{j}_{\bar{n}}^S&=-it(e^{-i\theta_N}\hat{c}^\dagger_{n+1}\hat{c}_{n}-e^{i\theta_N}\hat{c}^\dagger_{n}\hat{c}_{n+1})
\end{align}
with $t=J/2$ and $\hat{c}_{n+L}=\hat{c}_{n}$. For $\hat{\Gamma}\equiv\sum_{n=1}^L\hat{j}_{\bar{n}}^E$, operators $\hat{\xi}$ and $\hat{\zeta}$ read
\begin{align}
\hat{\xi}&=-\frac{1}{L}\sum_{n=1}^L2t^2(e^{-2i\theta_N}\hat{c}^\dagger_{n+2}\hat{c}_{n}+e^{2i\theta_N}\hat{c}^\dagger_{n}\hat{c}_{n+2}),\\
\hat{\zeta}&=-\frac{4}{L}\sum_{n=1}^L\hat{j}_{\bar{n}}^E=-\frac{4\hat{\Gamma}}{L}.
\end{align}

At the zero temperature, all single particle states with 
\begin{align}
\varepsilon_k-\mu-\eta \varepsilon_kv_k<0
\end{align}
are occupied and the expectation values are given as the sum over all occupied states:
\begin{align}
\langle\hat{s}^z\rangle^{(\mu,\eta)}&=-\frac{1}{2}+\frac{1}{L}\sum_{k:\text{occ}}1,\\
\langle\hat{j}^S\rangle^{(\mu,\eta)}&=\frac{1}{L}\sum_{k:\text{occ}}v_k,\\
\langle\hat{j}^E\rangle^{(\mu,\eta)}&=\frac{1}{L}\sum_{k:\text{occ}}\varepsilon_kv_k,\\
\langle\hat{\xi}\rangle^{(\mu,\eta)}&\equiv\text{tr}[\hat{\rho}_{\text{gG}}^{(\mu,\eta)}\hat{\xi}]=\frac{1}{L}\sum_{k:\text{occ}}\partial_k(\varepsilon_kv_k).
\end{align}
Here $\varepsilon_k=2t\cos(k+\theta_N)$ is the band dispersion and $v_k=\partial_k\varepsilon_k$ is the group velocity.

\begin{figure}
\begin{center}
\includegraphics[width=\columnwidth]{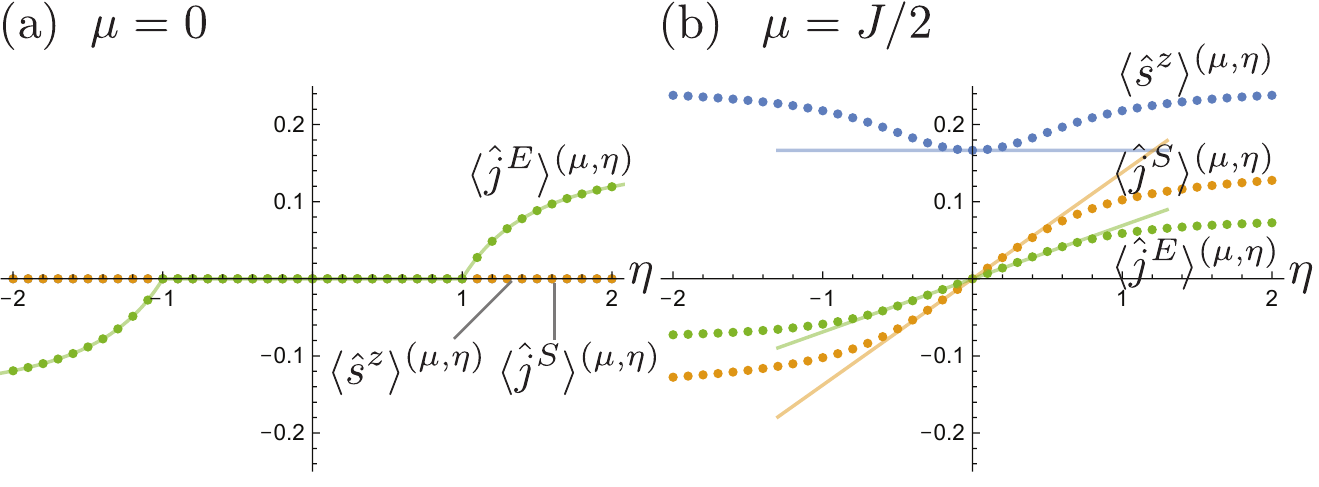}
\caption{\label{fig1} 
Numerical results of $\langle\hat{s}^z\rangle^{(\mu,\eta)}$ (blue dots), $\langle\hat{j}^S\rangle^{(\mu,\eta)}$ (orange dots), and $\langle\hat{j}^E\rangle^{(\mu,\eta)}$ (green dots) for (a) $\mu=0$ and (b) $\mu=J/2$. We set $J=1$ and $L=10^5\pm2$ in these plots. Solid curves represent analytic expressions for the thermodynamic limit in Eqs.~\eqref{ref1}--\eqref{ref4}.
}
\end{center}
\end{figure}

When $\mu=0$, Fermi points can be determined by the conditions $\epsilon_k=0$ or $1-\eta v_k=0$. In this case, we find the analytic expression for the large $L$ limit:
\begin{align}
\langle\hat{s}^z\rangle^{(0,\eta)}&=0,\\
\langle\hat{j}^S\rangle^{(0,\eta)}&=0,\\
\langle\hat{j}^E\rangle^{(0,\eta)}&=
\begin{cases}
\text{sign}\,\eta\,(J^2-\eta^{-2})/(2\pi)&(|\eta J|\geq1)\\
0&(|\eta J|<1)\\
\end{cases}\label{ref1},\\
\langle\hat{\xi}\rangle^{(0,\eta)}&=0.
\end{align}
When $\mu\neq0$, we find the following expansions for small $J\eta$:
\begin{align}
\langle\hat{s}^z\rangle^{(\mu,\eta)}&=\text{sign}J\,\left[\frac{1}{2}-\frac{\arccos(\mu/J)}{\pi}\right]+O((J\eta)^2),\\
\langle\hat{j}^S\rangle^{(\mu,\eta)}&=\text{sign}J\,\frac{\mu}{\pi}\left[\sqrt{1-(\mu/J)^2}J\eta+O((J\eta)^3)\right],\\
\langle\hat{j}^E\rangle^{(\mu,\eta)}&=\text{sign}J\,\frac{\mu^2}{\pi}\left[\sqrt{1-(\mu/J)^2}J\eta+O((J\eta)^3)\right],\label{ref4}\\
\langle\hat{\xi}\rangle^{(\mu,\eta)}&=|J|\frac{\mu}{\pi}\left[\sqrt{1-(\mu/J)^2}+O((J\eta)^2)\right].
\end{align}
These results are consistent with the general constraints in Eqs.~\eqref{c1}--\eqref{c3}.
In particular, we observe that $\langle\hat{j}^S\rangle^{(\mu,\eta)}=\eta \langle\hat{\xi}\rangle^{(\mu,\eta)}$ at least at the lowest order of $J\eta$, which is expected from Eq.~\eqref{Bloch32}.
We numerically demonstrate these results in Fig.~\ref{fig1}. 

\section{Conclusion}
In this work, we clarified the equivalences and the differences between different ensembles regarding the persistent current in non-equilibrium steady state. 
We confirmed that a nonzero persistent current in the limit of large system size can be supported in the canonical ensemble in the presence of an additional conserved charge $\hat{\Gamma}$, as in the generalized Gibbs ensemble~\cite{ElseSenthil}.
Although this was well-anticipated from the equivalence of different ensembles~\cite{landau}, the concrete derivation presented in Sec.~\ref{GCACC} uses the variational principle in a new way. 

There are also subtle differences related to the property of the reservoir. When the system and the reservoir exchange the charge $\hat{\Gamma}$, the corresponding ``chemical potential" $\eta$ must be balanced between the system and the reservoir after equibliration. Then, if the system is expected to support a nonzero persistent current, so is the reservoir. If the reservoir is an external system, the persistent current of the system vanishes in the co-moving frame of the reservoir.  In contrast, when the charge $\hat{\Gamma}$ is strictly conserved within the system, a persistent current in the large $L$ limit is allowed, in principle. 

\begin{acknowledgments}
We thank Takahiro Sagawa, Tatsuhiko N. Ikeda, and Yohei Fuji for useful discussions. 
The work of H. W. is supported by JSPS KAKENHI Grant No.~JP20H01825 and by JST PRESTO Grant No.~JPMJPR18LA. 
\end{acknowledgments}

\bibliography{GGEref.bib}
\end{document}